\documentclass[12pt]{article}
\newlength{\mytopmargin}
\newlength{\myleftmargin}
\newtheorem{theorem}{Theorem}[section]
\newtheorem{proposition}[theorem]{Proposition}

\usepackage{amsmath,amsfonts,amssymb}

\usepackage{graphicx}

\begin{document}

\title{Large deviation eigenvalue density for the soft edge Laguerre and Jacobi $\beta$-ensembles}
\author{Peter J. Forrester}
\date{}
\maketitle

\noindent
\thanks{\small Department of Mathematics and Statistics, 
The University of Melbourne,
Victoria 3010, Australia email: p.forrester@ms.unimelb.edu.au
}

\begin{abstract}
\noindent We analyze the eigenvalue density for the Laguerre and Jacobi $\beta$-ensembles in the cases that the
corresponding exponents are extensive. In particular, we obtain the asymptotic expansion up to terms $o(1)$, in the large
deviation regime outside the limiting interval of support. As found in recent studies of the large deviation density for the
Gaussian $\beta$-ensemble, and Laguerre $\beta$-ensemble with fixed exponent, there is a scaling from this
asymptotic expansion to the right tail asymptotics for the distribution of the largest eigenvalue at the soft edge.
\end{abstract}

\section{Introduction}

Let $X$ be a $p\times N$ ($p\geq N$) random matrix with standard real Gaussian entries. It is well known in random matrix theory \cite{Mu82, Fo10} that the eigenvalue probability density function (PDF) of the corresponding covariance matrix $W=X^TX$ is proportional to
\begin{align}
\label{1} \prod_{l=1}^{N} \lambda_l^{\beta a/2} e^{-\beta \lambda_l/2} \prod_{1\leq j < k \leq N} |\lambda_k -\lambda_j|^{\beta},
\end{align}
with $a=p-N+1-2/\beta$ and $\beta=1$. For general $\beta$ (\ref{1}) is referred to as the Laguerre $\beta$-ensemble and is to be denoted ${\rm ME}_{\beta, N} (\lambda^{\beta a/2} e^{-\beta\lambda/2})$.

In multivariate statistics, thinking of $X$ as a data matrix with successive columns storing the values of $p$ measurements for each of the $N$ observables,
knowledge of (\ref{1}) allows for an analysis of questions such as ``are the measured quantities in $X$ correlated?'' \cite{Mu82}. In physics, the Laguerre $\beta$-ensemble (\ref{1}) with exponent $N/2$ appears as the eigenvalue PDF of the Wigner-Smith delay time matrix in mesoscopic transport problems \cite{BFB96} (see also \cite[\S 3.3.1]{Fo10}). The parameter $\beta$ can then take on one of the three values $\beta=1,2$ or $4$ depending on the underlying time reversal symmetry. We remark too that the case $\beta=2$ of (\ref{1}) corresponds to the eigenvalue PDF for the so-called Penner model in the theory of matrix models (see 
e.g.~\cite{PS10}). It also appears in the analysis of information capacity in the setting of wireless communication systems \cite{TV04}.

A peculiar feature of the Wigner-Smith time delay matrix example is that the parameter $a$ in (\ref{1}) is extensive, being proportional to the number of eigenvalues $N$. This is in fact what is required in the analysis of the Penner model, and this same limit is of interest in the study of wireless communication systems; see references as cited above. One direction of study relating to (\ref{1}) with $a$ extensive is the large $N$ asymptotic expansion of the moments
\begin{align}
\label{2.1} \left\langle \sum_{l=1}^{N} \lambda_l^k\right\rangle_{{\rm ME}_{\beta, N}(\lambda^{\beta a/2} e^{-\beta\lambda/2})}.
\end{align}
Recent references on this topic include \cite{LV11}, \cite{MS11a, MS11b}. These moments probe the non-oscillatory portion of the large $N$ asymptotic expansion of the eigenvalue density in the domain of its leading support. Our interest in this paper is also in the large $N$ expansion of the eigenvalue density with $a$ extensive, but in the domain outside rather than inside its leading support.

Physical motivation to pursue this problem comes from a very recent  \cite{FPNFD11} experimental set up relating to (\ref{1}) in the case $\beta=2$, which in turn comes about by (\ref{1}) corresponding to the eigenvalue PDF for matrices $X^{\dag}X$, where $X$ is a $p\times N$ complex Gaussian matrix. In fact, the theoretical discussion in \cite{FPNFD11} tells us that the so-called round trip propagation matrix $M$ determining the evolution of the complex electric field for a $p\times N$ coupled array of high gain lasers can be decomposed in the form $X^{\dag}X$ for $X$ a complex $p\times N$ matrix with approximately Gaussian entries. Furthermore, the output power is proportional to the largest eigenvalue of $M$. With hundreds of thousands of data points accessible in this way, it is possible to probe the large deviation portions of the largest eigenvalue PDF from these measurements, something that was done explicitly for the square case $p=N (=5)$. For the right tail the large deviation form of the largest eigenvalue PDF is, to leading order, equal to the density. This happens in the domain outside the leading order support, where the density takes on exponentially small values.

There is another random matrix ensemble that allows for extensive exponents. This is the Jacobi $\beta$-ensemble, specified by an eigenvalue PDF proportional to
\begin{align}
\label{3.1} \prod_{l=1}^N \lambda_l^{\beta a_1/2} (1-\lambda_l)^{\beta a_2/2} \prod_{1\leq j < k\leq N} |\lambda_k - \lambda_j|^{\beta}, \qquad 0< \lambda_l< 1
\end{align}
and referred to as ${\rm ME}_{\beta, N} (\lambda^{\beta a_1/2} (1-\lambda^{\beta a_2/2}))$. In the special case $a_2=0, a_1= p-N+1-2/\beta$ and $\beta=1,2$ and $4$ (\ref{3.1}) gives the PDF of the non-zero squared singular values of an $N\times p$ sub-block of a $(N+p)\times (N+p)$ matrix from the circular orthogonal ensemble ($\beta=1$), circular unitary ensemble ($\beta=2$) and circular symplectic ensemble ($\beta=4$) (see \cite[\S 3.8.2]{Fo10}). The propagation of electron fluxes in a disordered mesoscopic wire is well described by scattering matrices from the circular ensembles. And an $N\times p$ sub-block then corresponds to the transmission matrix relating a state of $p$ channels entering the left of the wire and a state of $N$ channels exiting the right. The corresponding singular values then determine the conductance (see e.g. \cite{Be97}). We will compute the asymptotic form of the spectral density for (\ref{3.1}) with $a_1$ and $a_2$ extensive, in the asymptotically small region outside the leading support. With (\ref{3.1}) in the case $\beta=1$ occurring in mathematical statistics, and the largest eigenvalue having special significance in various statistical tests \cite{Jo08}, the fact that our results imply the right tail asymptotic form of the distribution of the largest eigenvalue thus provides us with  motivation in an applied setting.

Our task then is to compute the eigenvalue density for both the Laguerre and Jacobi $\beta$-ensembles with extensive exponents, in the exponentially small regime outside the leading support. 
In addition to the interest from the viewpoint of the problems discussed above, this complements the recent computation of the large deviation density
for the Gaussian $\beta$-ensemble \cite{Fo11,BN11}, and the Laguerre $\beta$-ensemble with $a$ fixed \cite{Fo11}. Here we will use the method of
\cite{Fo11} based on writing the density as an average of the $\beta$ moment of the characteristic polynomial, and then using
a Gaussian fluctuation theorem to deduce the asymptotics. This is distinct from the method of \cite{BN11} which is based on the loop equations for
correlators.
In Section 2 we review results on this problem already documented in the literature, and then present our general strategy. The details of our calculations, leading to the sought large deviation formulas, are given in Section 3. In the case of 
the Laguerre ensemble with weight $\lambda^{\beta \alpha N/2 + \beta/2 - 1} e^{-\beta \lambda /2}$, where $u^{\rm L}$ is given by (\ref{u}) below and
$x \in (0,( \sqrt{\alpha+1}-1)^2 ) \cup (( \sqrt{\alpha+1}+1)^2, \infty)$, we obtain
\begin{align}\label{A1}
N\rho_{(1),N}^{\rm L} (Nx) \mathop{\sim}\limits_{N\to\infty} & \exp \Bigg\{ \frac{N\beta}{2}\Bigg ( -u^{\rm L} + 2 \log \left| \frac{u^{\rm L} +x-2-
\alpha}{2(1+\alpha)^{1/2}}\right|  
-2 \alpha\log \left| \frac{u^{\rm L}  -x-\alpha}{2x^{1/2}(1+\alpha)^{1/2}} \right| \Bigg ) \Bigg \} \nonumber \\
\nonumber & \times \exp \Bigg \{  (1-3\beta/2) \log  |u^{\rm L} |  - (1 - \beta/2) 
\log \left| \frac{x(u^{\rm L} +x-2-
\alpha)}{2}\right| \Bigg \}
\\
&\times  \frac{N(1+\alpha)^{1/2}}{2\pi} \left( \frac{2}{\beta N}\right)^{\beta/2} \Gamma (1+\beta/2)
\end{align}
while for the  Jacobi ensemble with weight $\lambda^{\beta \alpha_1 N/2 +\beta/2 - 1}(1 - \lambda)^{\beta \alpha_2 N/2 +\beta/2 - 1}$, where
$u^{\rm J}$ is given by (\ref{17.bx}) below and $x \in (0,c_1) \cup (c_2,1)$ with $c_1, c_2$ given by (\ref{cd1}), (\ref{cd2}), our final result reads
\begin{align}\label{A2}
\rho_{(1),N}^{\rm J} (x) \mathop{\sim}\limits_{N\to\infty} & \exp \Bigg\{  N \beta \Bigg ( \log \Bigg |
{x - (c_1 + c_2)/2 + u^{\rm J} \over (c_2 - c_1)/2} \Bigg | \nonumber \\
& - \alpha_1 \log {| \sqrt{c_1 c_2} + x - u^{\rm J} | \over (\sqrt{c_1} + \sqrt{c_2} )x^{1/2}} -
 \alpha_2 \log {| \sqrt{(1 -c_1)(1 - c_2)} -  x + 1 + u^{\rm J} | \over
 (\sqrt{1 - c_1} + \sqrt{1 - c_2}) (1 - x)^{1/2}} \Bigg ) \Bigg \} \nonumber \\
 &  \times \exp \Bigg \{  (1-3\beta/2) \log  |u^{\rm J} | - (1 - \beta/2) 
\log{ \left| x(1-x)(u^{\rm J} +x- (c_1 + c_2)/2) \over 2 + \alpha_1 + \alpha_2
\right| }\Bigg \}  \nonumber \\
& \times   \frac{N(c_2 - c_1)}{4 \pi } \left( \frac{1}{\beta N}\right)^{\beta/2} \Gamma (1+\beta/2).
\end{align} 
The notation $\sim$ in (\ref{A1}) and (\ref{A2}) means that LHS/RHS $\to 1$ in the appropriate limit.
Both these asymptotic formulas are shown to exhibit a scaling to the asymptotic form of the right tail for the distribution of the largest eigenvalue
at the soft edge. 
In the Appendix we critique a recent \cite{KC10} claimed large deviation-type asymptotic expansion for the smallest eigenvalue PDF in the Laguerre $\beta$-ensemble in the case that $a$ is fixed. 

\section{Previous results and strategy}
\setcounter{equation}{0}
\subsection{Previous results}

For the Laguerre $\beta$-ensemble with $a=\alpha N + O(1)$ the leading support of the eigenvalue density, upon scaling the eigenvalues $\lambda_l\mapsto N\lambda_l$, is the interval $(a_1^2, a_2^2)$ with
\begin{align}
\label{cb} a_1 = \sqrt{\alpha+1}-1, \qquad \quad a_2=\sqrt{\alpha+1}+1. 
\end{align}
Moreover, the corresponding leading eigenvalue density is given by the so-called Mar\v{c}enko-Pastur law
\begin{align}
\label{3.a} N\rho_{(1)} (Nx) \sim \frac{N}{2\pi x} \sqrt{\left( a_2^2-x \right) \left(x-a_1^2 \right)}, \qquad a_1^2<x<a_2^2
\end{align}
(see e.g. \cite[eq. (3.61) with $x^2\mapsto x$]{Fo10}). Thus for $\alpha >0$ the smallest eigenvalue is bounded away from the origin with a density profile that vanishes like a square root (a soft edge), while for $\alpha=0$ the smallest eigenvalue is part of an accumulation of eigenvalues about the origin with a density profile diverging like a reciprocal of a square root (a hard edge).

A similar effect is exhibited by the Jacobi $\beta$-ensemble (\ref{3.1}) with $a_1=\alpha_1 N + O(1), a_2=\alpha_2 N + O(1)$, except now there is no need to scale the eigenvalues. Thus the leading eigenvalue support is $(c_1, c_2)\subset [0,1]$ with
\begin{align}
\label{cd1} 2(c_1+c_2-1) &= \frac{2(\alpha_1^2 -\alpha_2^2)}{(\alpha_1+\alpha_2+2)^2}\\
\label{cd2} (2c_1-1)(2c_2-1) &= \frac{2(\alpha_1^2+\alpha_2^2)}{(\alpha_1+\alpha_2+2)^2}-1.
\end{align}
Notice that (\ref{cd1}) and (\ref{cd2}) are unchanged by
\begin{align}
\label{map} (c_1, c_2, \alpha_1, \alpha_2)\mapsto (1-c_1, 1-c_2, \alpha_2, \alpha_1).
\end{align}
The corresponding eigenvalue density is given by
\begin{align}
\label{cd3} \rho_{(1)} (x) \sim N\frac{2+\alpha_1+\alpha_2}{2\pi} \frac{\sqrt{(x-c_1)(c_2-x)}}{x(1-x)}
\end{align}
(see e.g. \cite[eq. (3.77) with $y\mapsto 2y-1$]{Fo10}).

We are interested in the eigenvalue density outside of $\left[ a_1^2, a_2^2\right]$ for the Laguerre $\beta$-ensemble, and outside of $[c_1, c_2]$ for the Jacobi $\beta$-ensemble. In the former case there are some prior results from the existing literature, which we will now summarize.
But
before doing so, we make note of the work \cite{MV09}. This
 relates to the case $\alpha=0$ in the Laguerre ensemble, and so strictly speaking is not relevant to our setting of extensive exponents. However, the leading term of the asymptotic expansion of the density does not contain any information on the exponent $a$ for $a$ of order unity, and so is expected to
 correspond to the case $\alpha=0$. Actually the concern of \cite{MV09} is not the density in the exponentially small region as such, but the large deviation asymptotic form of the distribution of the largest eigenvalue $p_{N,\beta}^{\rm L}(s)$. However, it is a simple result that for $x$ in this regime, the two are identical \cite[eq. (14.136)]{Fo10}, 
\begin{align}
\label{11.1} p_{N, \beta}^{\rm L}  (Nx) \sim \rho_{(1)}^{\rm L}(Nx).
\end{align}
The leading large $N$ form of (\ref{11.1}) for $x>4$ in the case $\alpha=0$ as calculated in \cite{MV09} is given by setting $\alpha=0$ in (\ref{12.1}) below.

There are at least three existing works containing explicit asymptotic formulas for
$\rho_{(1)}^{\rm L} (Nx)$ (or equivalently, according to (\ref{11.1}), $p_{N, \beta}^{\rm L}(Nx)$) outside of $\left[a_1^2, a_2^2 \right]$
for general $\alpha$. In order of publication date the first \cite[eq. (4.83)]{Fo10} is the leading large $N$ form
\begin{align}
\label{12.1} \rho_{(1)}^{\rm L} &(Nx) \nonumber \\ &
 = \exp \left \{ \frac{\beta N}{2} \left( -u^{\rm L} +\alpha \log \left| \frac{\alpha(\alpha +u^{\rm L}-x) -2x} {2(\alpha+1)^{1/2} x} \right|+(2+\alpha) \log \left| \frac{u^{\rm L}+x-2-\alpha}{2(\alpha+1)^{1/2}} \right|\right) + o(N)\right \} ,
\end{align}
where
\begin{align}
\label{u} u^{\rm L} := \left \{
\begin{array}{ll} \displaystyle \sqrt{(x - a_1^2)(x - a_2^2)}, & x \in (a_2^2,\infty) \\  \displaystyle
-\sqrt{(x - a_1^2)(x - a_2^2)}, & x \in (0,a_1^2).\end{array}  \right. 
\end{align}
The second \cite{KC10} is presented as the leading asymptotic form for $p_{N, \beta}^{\rm L}(Nx)$ in the region $x\in [0, a_1^2]$. In terms of our notation the result of \cite{KC10} reads
\begin{align}
\label{12.2} \rho_{(1)}^{\rm L} (Nx) =  \exp \left( -\beta N \Phi^{\min}_{-} \left( (1-\sqrt{\alpha +1})^2-x\right) + o(N) \right),
\end{align}
where
\begin{align}
\nonumber \Phi_{-}^{\min} (x) &= -\frac{1}{2}\sqrt{x\left(x+4\sqrt{\alpha +1}\right)}- \frac{\alpha} {2} \log \left( 1-\frac{x}{(1-\sqrt{\alpha+1})^2} \right)\\
\nonumber &+2\log \left( \frac{\sqrt{x+4\sqrt{\alpha+1}}-\sqrt{x}} {\sqrt{4\sqrt{\alpha+1}}} \right)\\
\label{13.1} &+\alpha \log \left( 1+ \frac{2\sqrt{x}}{\sqrt{\alpha+1}-1} \left( \frac{\sqrt{x+4\sqrt{\alpha+1}}-\sqrt{x}} {4\sqrt{\alpha+1}}\right) \right).
\end{align}
Elementary algebraic manipulation involving the simple identities
\begin{eqnarray}
\lefteqn{
\alpha \log \Bigg | {\alpha(\alpha + u^{\rm L} - x) - 2x \over 2 (\alpha + 1)^{1/2}} \Bigg | +
\alpha \log  \Bigg |  {u^{\rm L} + x - 2 - \alpha \over 2 (\alpha + 1)^{1/2} x} \Bigg | }
\nonumber \\
& \qquad = \displaystyle
2 \alpha \log \Bigg | {\alpha + u^{\rm L}  +x \over 2 \sqrt{\alpha + 1} \sqrt{x}} \Bigg | =
-2 \alpha \log \Bigg | {\alpha - u^{\rm L}  + x \over 2 \sqrt{\alpha + 1} \sqrt{x}} \Bigg |
\end{eqnarray}
shows that (\ref{12.1}) and (\ref{12.2}) are in fact identical expressions.

The third and most recent result \cite{ATK11} was derived for the case $\beta=2$ only, but in addition to the leading exponential term also gives the algebraic correction. Thus, after making the correspondences $\lambda=\alpha, \xi_0 = a_1^2/(\alpha+1), \xi_1=a_2^2/(\alpha+1), x\mapsto x/(\alpha+1)$ we read off from \cite{ATK11} that for exponent $a = \alpha N$
\begin{align}
\label{14.1} \rho_{(1)}^{\rm L} \big(Nx\big) \sim \frac{ (\alpha+1)^{1/2}} {2\pi N} \frac{1}{(x-a_1^2) (x-a_2^2)} e^{-N \kappa(x)},
\end{align}
where
\begin{align}\label{14.1a}
 \kappa(x) = - u^{\rm L}  + a_1a_2\: \mathrm{arcosh} \left| \frac{\frac{1}{x} - \frac{1}{2} \left( \frac{1}{a_1^2} + \frac{1}{a_2^2}\right)} {\frac{1}{2} \left( \frac{1}{a_1^2}- \frac{1}{a_2^2} \right)} \right|
 - \frac{a_1^2+a_2^2}{2} \: \mathrm{arcosh} \left| \frac{\frac{a_1^2+ a_2^2}{2} -x}{\frac{a_1^2- a_2^2}{2}}\right|.
\end{align}
Recalling that for $x>1$
\begin{align}
\nonumber \mathrm{arcosh} x = \log (x+\sqrt{x^2-1})
\end{align}
shows that the leading (exponential) term of (\ref{14.1}) coincides with the $\beta=2$ case of (\ref{12.1}).

\subsection{Strategy}

We seek to extend (\ref{12.1}) to the same order as the $\beta=2$ result (\ref{14.1}), and thus give the asymptotic form of $\rho_{(1)}^L (Nx)$ for $x\in [0, a_1^2)\cup (a_2^2, \infty )$ up to terms $o(1)$ as $N\to \infty$. We seek too an analogous expansion in the Jacobi case. Our method relies on the fact that the density can be written in terms of the $\beta$-moment of the corresponding characteristic polynomial, and as such is an application of the strategy used recently in
the case of the Gaussian $\beta$-ensemble, and the Laguerre $\beta$-ensemble with $a$ fixed \cite{Fo11}.

We have referred to the PDFs (\ref{1}) and (\ref{3.1}) by ${\rm ME}_{\beta, N}(w)$ with $w(\lambda) = \lambda^{a\beta/2} e^{-\beta \lambda/2}$ and $w(\lambda)= \lambda^{a_1\beta/2} (1-\lambda)^{a_2\beta/2}$ respectively. Let us denote the corresponding normalizations by $C_{\beta, N}[w]$. Then it follows from the definitions (see e.g. \cite[eq. (2.11)]{Fo11}) that 
\begin{align}
\label{2.11} \rho_{(1), N+1} (x)= \frac{(N+1) C_{\beta, N}[w]} {C_{\beta, N+1}[w]} w(x) \left\langle \prod_{l=1}^N |x-\lambda_l|^{\beta} \right\rangle_{{\rm ME}_{\beta, N}(w)}.
\end{align}
The key observation is that the average in (\ref{2.11}) can be interpreted as a particular case of the characteristic function for the linear statistic $V(x)= \sum_{l=1}^N \log |x-\lambda_l|$. But for a large class of weights $w$ the latter is expected to be a Gaussian
(see e.g.~\cite[\S 14.4]{Fo10}), and this implies, with $x$ outside the interval of support, 
\begin{align}
\label{17.1} \left\langle \prod_{l=1}^N |x-\lambda_l|^{\beta} \right\rangle_{{\rm ME}_{\beta, N}(w)} \sim e^{\beta \mu_N (v)} e^{(\beta \sigma (v))^2/2},
\end{align}
where, with $v(t):= \log |x-t|$ and $(d_1, d_2)$ the limiting interval of support, 
\begin{align}
\label{17.2a} \mu_N(v) &= \int_{d1}^{d_2} \rho_{(1), N}(t) v(t) dt\\
\nonumber (\sigma(v))^2 &= \frac{1}{\beta \pi^2} \int_{d_1}^{d_2} dt_1 \frac{v(t_1)} {\big((d_2-t_1)(t_1-d_1)\big)^{1/2}} \int_{d_1}^{d_2} dt_2 \frac{v'(t_2) \big((d_2-t_2) (t_2-d_1)\big)^{1/2}} {t_2-t_1}\\
\label{17.2b} &= \frac{1}{2\beta} \sum_{k=1}^{\infty} k a^2_k, \qquad a_k=\frac{2}{\pi}\int_0^{\pi} v\left( \frac{d_1+d_2}{2} + \frac{d_2-d_1}{2}\cos \theta \right) \cos k\theta \; d\theta.
\end{align}

Theorems relating to the asymptotic form of the characteristic function for linear statistics in random matrix theory were first proved in \cite{Jo98}. However, the technical assumptions therein did not permit the linear statistic corresponding to $v(t) = \log |x-t|$ due to its singularity at $x=t$. But for $x$ outside the eigenvalue support (\ref{17.2b}) is well defined and on physical grounds it is expected that (\ref{17.1}) will still apply \cite[\S 14.4]{Fo10}. Fortunately, very recently \cite{BG11}, under the conditions that the eigenvalue support is a single interval, and with both endpoints soft edges (or a hard edge with parameter $a=0$) the validity of (\ref{17.1}) for any $v$ analytic in the neighbourhood of $[d_1, d_2]$ has been rigorously established. In our setting of the Laguerre and Jacobi ensembles with extensive parameters both end points of the eigenvalue support are soft edges, so (\ref{17.1}) thus follows from \cite{BG11}.

Our immediate task then is to compute the explicit form of (\ref{17.2a}) and (\ref{17.2b}). We will perform this task in the next section. Substituting (\ref{17.1}) in (\ref{2.11}) then gives the sought asymptotic expansions of the densities.

At a technical level, with the parameters being extensive, and the occurrence of both $N+1$ and $N$ in (\ref{2.11}), there is some advantage in first manipulating (\ref{2.11}) before applying (\ref{17.1}). Thus in the Laguerre case the weight required in (\ref{2.11}) is
\begin{align}
\label{2.11'} w(x) = x^{\alpha (N+1)\beta/2 + \beta/2 - 1} e^{-(N+1)\beta x/2}.
\end{align}
With this $w$ we observe
\begin{align}
\label{2.11x} \left\langle \prod_{l=1}^N |x-\lambda_l|^{\beta}\right\rangle_{{\rm ME}_{\beta, N}(w)} = 
{C_{\beta, N}[\tilde{w}]  \over C_{\beta, N}[w] }
 \left\langle \prod_{l=1}^N \lambda_l^{\alpha \beta/2} e^{-\beta\lambda_l/2}|x-\lambda_l|^{\beta}\right\rangle_{{\rm ME}_{\beta, N}(\tilde{w})},
\end{align}
where $\tilde{w}(x) =x^{\alpha N\beta/2 + \beta/2 - 1} e^{-N\beta x/2}$, and we observe too that
\begin{equation}\label{2.20a}
{C_{\beta, N}[\tilde{w}]  \over C_{\beta, N}[w] } = {1 \over \langle \prod_{l=1}^N \lambda_l^{\alpha \beta /2} e^{-\beta \lambda_l/2}
\rangle_{{\rm ME}_{\beta, N}(\tilde{w})}}.
\end{equation}

Similarly, in the Jacobi case, the weight required in (\ref{2.11}) is
\begin{align}\
\label{2.11g} w(x) = x^{\alpha_1 (N+1)\beta/2 +\beta/2 -1} (1-x)^{\alpha_2(N+1) \beta /2  +\beta/2 -1}
\end{align}
and we have 
\begin{align}
\label{2.11y} \left\langle \prod_{l=1}^N |x-\lambda_l|^{\beta}\right\rangle_{{\rm ME}_{\beta, N}(w)} = 
{C_{\beta, N}[\tilde{w}] \over C_{\beta, N}[w]  }
\left\langle \prod_{l=1}^N \lambda_l^{\alpha_1 \beta/2} (1-\lambda_l)^{\alpha_2 \beta /2}|x-\lambda_l|^{\beta}\right\rangle_{{\rm ME}_{\beta, N}(\tilde{w})},
\end{align}
where $\tilde{w}(x) =x^{\alpha_1 N\beta/2 +\beta/2 - 1} (1-x)^{\alpha_2 N\beta/2 +\beta/2 - 1}$,
together with
\begin{equation}\label{2.22a}
{C_{\beta, N}[\tilde{w}]  \over C_{\beta, N}[w] } = {1 \over \langle \prod_{l=1}^N \lambda_l^{\alpha_1 \beta /2} (1 - \lambda_l)^{\alpha_2 \beta/2}
\rangle_{{\rm ME}_{\beta, N}(\tilde{w})}}.
\end{equation}

 Our immediate task then is to compute the explicit form of (\ref{17.2a}) and (\ref{17.2b}) for the averages on the RHS's of (\ref{2.11x}) and (\ref{2.11y}) with
\begin{align}
\label{vLJ} v(t)= \left\{\begin{array}{ll}
\log |x-t| +\frac{\alpha}{2} \log t - \frac{1}{2}t, & \mathrm{Laguerre}\\
\log |x-t| +\frac{\alpha_1}{2} \log t +\frac{\alpha_2}{2} \log |1-t|, & \mathrm{Jacobi},
\end{array}\right.
\end{align}
and with
\begin{align}
\label{2.23a} v(t)= \left\{\begin{array}{ll}
\frac{\alpha}{2} \log t - \frac{1}{2}t, & \mathrm{Laguerre}\\
\frac{\alpha_1}{2} \log t +\frac{\alpha_2}{2} \log |1-t|, & \mathrm{Jacobi}.
\end{array}\right.
\end{align}
For future purposes, especially in labellings, it will be convenient to refer to (\ref{vLJ}) and (\ref{2.23a}) as choices (1) and (2)
respectively. We will proceed with required computations in the next section.

\section{The asymptotic forms}
\setcounter{equation}{0}
\subsection{The Laguerre $\beta$-ensemble}

For the weight $\tilde{w}$ in (\ref{2.11x}) it is known from \cite{DE05} that
\begin{align}
\nonumber \rho_{(1), N}(t) = &\frac{N}{2\pi t} \sqrt{\left(a_2^2- t \right)\left( t-a_1^2\right)}\\
\label{12.a} & +\left( \frac{1}{\beta} - \frac{1}{2}\right) \left( \frac{1}{2} \delta(t-a_1^2) +\frac{1}{2} \delta (t-a_2^2) - \frac{1}{\pi} \frac{1}{\sqrt{(t-a_1^2) (a_2^2-t)}}\right)
+ O \Big ( {1 \over N} \Big ).
\end{align}
We remark that the precise form of the $O(1)$ term in (\ref{12.a}) relies crucially on the $O(1)$ portion of the exponent in $\tilde{w}$
equalling $\beta/2 - 1$. Note that the latter is precisely the $O(1)$ portion of the exponent appearing in (\ref{1}) in relation to covariance matrices.
We remark too that the expansion (\ref{12.a}) ignores possible oscillatory terms which appear at next order \cite{DF06}, as these do not contribute to the next correction in (\ref{17.2a}).

As is consistent with the labelling noted below (\ref{2.23a}), let  (\ref{17.2a}) in the case $\rho_{(1), N}(t)$ is given by (\ref{12.a}), and $v(t)$ by the
Laguerre cases of (\ref{vLJ}) and (\ref{2.23a}), be denoted $\mu_N^{L,(1)}$ and $\mu_N^{L,(2)}$ respectively. Then $\mu_N^{{\rm L},(1)} -
\mu_N^{{\rm L},(2)}$ is equal to (\ref{17.2a}) with $\rho_{(1), N}(t)$ given by (\ref{12.a}) and $v(t) = \log | x - t|$. It can be evaluated by making use of
the following integral evaluations.

\begin{proposition}
\label{L1}
Let $u^{\rm L}$ be given by (\ref{u}). For $x\in (0,a_1^2)\cup (a_2^2,\infty)$ we have
\begin{align}
\nonumber &\int_{a_1^2}^{a_2^2} \frac{1}{2t\pi} \sqrt{(t-a_1^2)(a_2^2-t)} \, \log |x-t| \, dt = \\
\label{20.1} & \frac{1}{2}\left( x-\alpha-u^{\rm L}-2+\alpha \log \left| \frac{\alpha(\alpha +u^{\rm L}-x)-2x}{2x^2}\right| + (2+\alpha) \log \left| \frac{u^{\rm L}+x-2-\alpha}{2}\right|\right)
\end{align}
and 
\begin{align}
\nonumber {1 \over \pi} \int_{a_1^2}^{a_2^2} \frac{\log |x-t|}{\sqrt{(a_2^2-t)(t-a_1^2)}} \,dt &= \log \frac{a_2^2-a_1^2}{2} +\log \frac{1}{2} \left|\tilde{x} \pm \sqrt{\tilde{x}^2-1} \right|\\
\label{20.2} &= \log \left|\frac{u^{\rm L}+x-2-\alpha}{2}\right|,
\end{align}
where in the first equality of (\ref{20.2}) we take $+\; (-)$ according to $x\in (a_2^2,\infty)\; \big(x\in (0,a_1^2) \big)$ and
\begin{align}
\label{xt} \tilde{x} = \frac{2}{a_2^2-a_1^2} \left( x- \frac{a_1^2+a_2^2}{2}\right).
\end{align}
%Also, it follows from (\ref{20.1}) that
%\begin{align}
%\label{21.d} \int_{a_1^2}^{a_2^2} \frac{1}{2\pi t} \sqrt{(t-a_1^2)(a_2^2 -t)} \log t \; dt &= - (1+\alpha) +(1+\alpha)\log (1+\alpha),\\
%\label{21.e} \int_{a_1^2}^{a_2^2} \frac{1}{2\pi} \sqrt{(t-a_1^2)(a_2^2 -t)}\; dt & = \alpha +1.
%\end{align}
\end{proposition}

\noindent
\textit{Proof}: The result (\ref{20.1}) is derived in \cite[Exercises 14.4 q.6(i)]{Fo10} for $\mathrm{Re} (x)>a_2^2$; its value for $x\in (0,a_1^2)$ follows by analytic continuation. The first equality in (\ref{20.2}) is a simple corollary of the integral evaluation
\begin{align}
\nonumber \frac{1}{\pi} \int_{-1}^{1} \frac{\log |x-t|}{\sqrt{1-t^2}} dt = \log \left( \frac{1}{2} (x+\sqrt{x^2-1})\right), \qquad x>1,
\end{align}
the derivation of which can be found, for example, in \cite[eq. (3.3)]{Fo11}. The second equality in (\ref{20.2}) now follows by noting
\begin{align}
\label{xua} \pm \frac{a_2^2-a_1^2}{2} \sqrt{\tilde{x}^2-1} = u^{\rm L}.
\end{align}
\hfill $\Box$

\medskip
%Turning our attention now to (\ref{21.d}) and (\ref{21.e}), we note from the definition (\ref{u}) that for $x\to 0$
%\begin{align}
%\nonumber u^{\rm L} = -\alpha \left( 1+\frac{\alpha+2}{\alpha^2}x +\frac{2(\alpha+1)}{\alpha^2}x^2+ \cdot\cdot\cdot \right)
%\end{align}
%while for $x\to \infty$
%\begin{align}
%\nonumber u^{\rm L} = x- (\alpha+2) -\frac{2(\alpha+1)}{x}+ \cdot\cdot\cdot
%\end{align}
%Using these expansions in (\ref{20.1}) allows the corresponding limits to be analyzed and implies (\ref{21.d}) and (\ref{21.e}) respectively. \hfill $\Box$

Proposition \ref{L1} implies that (\ref{17.2a}) in the case $\rho_{(1), N}(t)$ is given by (\ref{12.a}) and $v(t)$ by the Laguerre case of (\ref{vLJ}) has the 
large $N$ evaluation 
\begin{align}
\nonumber \mu_N^{{\rm L},(1)} &-
\mu_N^{{\rm L},(2)} = \frac{N}{2} \bigg ( x-\alpha -u^{\rm L}-2 +\alpha \log \left| \frac{\alpha(\alpha+u^{\rm L}-x)-2x}{2x^2}\right|\\
\label{11.4A} &  + (2+\alpha) \log 
\left| \frac{u^{\rm L}+x-2-\alpha}{2}\right|\bigg ) +\left( \frac{1}{\beta}- \frac{1}{2}\right) \left( \log |u^{\rm L}| -\log \left| \frac{u^{\rm L}+x-2-\alpha}{2} \right|\right) +
O \Big ( {1 \over N} \Big ).
\end{align}

It remains to evaluate (\ref{17.2b}) with $v(t)$ given by the Laguerre cases of (\ref{vLJ}) and (\ref{2.23a}). 
Consider the former case first. Making
use of the second equality in (\ref{17.2b}) shows we must evaluate
\begin{align}
\label{aa} a_k = a_k^{(1)} +a_k^{(2)} +a_k^{(3)}, \qquad k=1,2,...
\end{align}
where
\begin{align}
\label{I.a} a_k^{(1)} = \frac{2}{\pi} \int_0^{\pi} \log \left( 1-\frac{\cos \theta}{\tilde{x}} \right) \cos k\theta \; d\theta,
\end{align}
with $\tilde{x}$ given by (\ref{xt}), and
\begin{align}
\label{I.b} a_k^{(2)}&= \frac{\alpha}{2} a_k^{(1)} \Big|_{x=0},\\
\label{I.c} a_k^{(3)}&= - \frac{a_2^2-a_1^2}{2\pi} \int_0^{\pi} \cos \theta \cos k\theta \; d\theta.
\end{align}
From \cite[Lemma 2]{Fo11} we know that
\begin{align}
\label{I.d} \log \left( 1-\frac{\cos \theta}{\tilde{x}} \right) = \log (1-\nu_x e^{i\theta}) (1-\nu_x e^{-i\theta}) - \log (1+\nu_x^2),
\end{align}
where $\nu_x$ is such that $|\nu_x| <1$ and has the explicit value
\begin{align}
\label{nu} \nu_x = \left\{\begin{array}{cc}
\tilde{x}- (\tilde{x}^2-1)^{1/2}, & x\in (a_2^2, \infty),\\
\tilde{x}+ (\tilde{x}^2-1)^{1/2}, & x\in (0, a_1^2).
\end{array}\right.
\end{align}
We see immediately from (\ref{I.d}) that
\begin{align}
\label{nu1} a_k^{(1)} = - \frac{2 \nu_x^k}{k}.
\end{align}
According to (\ref{I.b}), we see from this, (\ref{nu}), (\ref{xt}) and (\ref{cb}) that
\begin{align}
\label{nu2} a_k^{(2)}= -\frac{\alpha}{k} \nu_0^k, \qquad \nu_0= -\frac{1}{\sqrt{\alpha+1}}.
\end{align}
And in relation to (\ref{I.c}), an elementary calculation shows
\begin{align}
\label{nu3} a_k^{(3)}= \left\{\begin{array}{cc}
-\sqrt{\alpha+1}, & k=1,\\
0, & \mathrm{otherwise}.
\end{array}\right.
\end{align}
Substituting these explicit forms in (\ref{aa}) we see
\begin{align}
\nonumber \sum_{k=1}^{\infty} k a_k^2 = &\alpha +1 +2\sqrt{\alpha+1}\left( 2\nu_x +\alpha \nu_0\right) + \sum_{k=1}^{\infty} \frac{1}{k} \left( 2\nu_x^k+ \alpha \nu_0^k\right)^2\\
\nonumber  = & -\alpha+1 +2 (x-(\alpha+2) -u^{\rm L}) -4\log (1-\nu_x^2)\\
\label{nu4} & - 4\alpha\log (1-\nu_x\nu_0) -\alpha^2\log (1-\nu_0^2).
\end{align}

With $v(t)$ given by the Laguerre case of (\ref{2.23a}), in the notation of (\ref{aa}) we have $a_k = a_k^{(2)} + a_k^{(3)}$.
The first line of (\ref{nu4}) then remains valid but with $\nu_x = 0$, so in this case we have
\begin{equation}\label{3.18a}
\sum_{k=1}^\infty k a_k^2 = - \alpha + 1 - \alpha^2 \log (1 - \nu_0^2).
\end{equation}

Recalling now  (\ref{17.2b}) and using a superscript notation analogous to that used in (\ref{11.4A}) we read off from (\ref{nu4}) and
(\ref{3.18a}) that
\begin{equation}\label{3.18b}
(\sigma^2)^{{\rm L},(1)} - (\sigma^2)^{{\rm L},(2)} = {1 \over 2 \beta} \Big ( 2 (x - (\alpha  + 2) - u^{\rm L}) -
4 \log (1 - \nu_x^2) - 4 \alpha \log (1 - \nu_x \nu_0) \Big ).
\end{equation}
Use of (\ref{nu}) and (\ref{nu2}) allows us to compute that
\begin{align}
\nonumber \log (1-\nu_x^2) &= \log |u^{\rm L}| - \log \frac{|x-(\alpha+2) +u^{\rm L}|}{2},\\
\log (1-\nu_x \nu_0) &= \log \frac{|x+\alpha-u^{\rm L}|}{2(\alpha+1)}. \label{nu5}
\end{align}
Substituting in (\ref{3.18b}) then gives the explicit evaluation of the variance
\begin{align}
\nonumber  (\sigma^2)^{{\rm L},(1)} - (\sigma^2)^{{\rm L},(2)}= &\frac{1}{\beta} (x-(\alpha+2) -u^{\rm L})-\frac{2}{\beta} \log |u^{\rm L}| + \frac{2}{\beta} \log \frac{|x-(\alpha+2) +u^{\rm L}|}{2}\\
\label{nu6} &-\frac{2\alpha}{\beta} \log \frac{|x+\alpha-u^{\rm L}|}{2(\alpha+1)} .
\end{align}

According to (\ref{17.1}) the asymptotic form of the product of the averages in (\ref{2.11x}) and (\ref{2.20a}) is determined,
up to terms vanishing as $N \to \infty$, by (\ref{11.4A}) and (\ref{3.18b}), allowing us to read off from these that
\begin{align}
\nonumber &\left\langle \prod_{l=1}^N |x-\lambda_l |^\beta \right\rangle_{{\rm ME}_{\beta, N}(\lambda^{\alpha(N+1)\beta/2 + \beta/2 - 1} 
e^{-(N+1)\beta \lambda/2})}\\
\nonumber & \mathop{\sim}\limits_{N\to\infty} \exp \left\{ \frac{(N+1)\beta}{2}\left( x-u^{\rm L}-\alpha-2+ 2\log \left| \frac{u^{\rm L}+x-2-\alpha}{2}\right| -2\alpha\log \left| \frac{u^{\rm L}-x-\alpha}{2(1+\alpha)}\right|\right) \right\}\\
\label{16.1} &\qquad\times \exp \left\{ (1-3\beta/2) \log |u^{\rm L}| - (1-\beta/2) \log \left| \frac{u^{\rm L}+x-2-\alpha}{2} \right|\right\}.
\end{align}
Although in the derivation $x$ has been assumed fixed, taking $x\to \infty$ shows that the RHS tends to $x^{N\beta}$, which is the leading form of the LHS. This suggests that in fact (\ref{16.1}) is valid uniformly for $x\in (a_2^2, \infty)$. We remark that the average of $\prod_{l=1}^N | x - \lambda_l |^\beta$ for the
Gaussian $\beta$-ensemble, and the Laguerre $\beta$-ensemble with exponent $a$ fixed, was shown to exhibit the same property \cite{Fo11}.

The announced  formula (\ref{A1}) follows by substituting (\ref{16.1}) in (\ref{2.11}) together with the asymptotic formula for the ratio of normalizations
\begin{align}
\nonumber \frac{C_{\beta, N}\left[ \lambda^{\alpha(N+1)\beta/2 +\beta/2 -1 } e^{-(N+1)\beta\lambda/2}\right]}{C_{\beta, N+1}\left[ \lambda^{\alpha(N+1)\beta/2
+\beta/2 -1} e^{-(N+1)\beta\lambda/2 +\beta/2 -1 }\right]} \sim \;& \frac{1}{2\pi} \left( \frac{2}{\beta N} \right)^{\beta/2} e^{\beta (N+1)(1+\alpha/2)}\\
\nonumber & \times\Gamma (1+\beta/2) (1+\alpha)^{-(N+1)(1+\alpha)\beta/2  + 1/2},
\end{align}
(a consequence of the explicit gamma function evaluation \cite[Prop. 4.7.3]{Fo10} and Stirling's formula).
This
formula  is in precise agreement with the result (\ref{14.1}) from \cite{ATK11}. 
With $\alpha = 0$ and $x$ scaled $x \mapsto 4x$ (\ref{A1}) reduces to the $a=1/2 - 1/\beta$ case of the asymptotic formula for the
density $\rho_{(1)}^{\rm L}(4Nx)$, $x > 1$, in the Laguerre ensemble with weight $w(\lambda) = \lambda^{\beta a/2} e^{-\beta \lambda / 2}$
given recently in \cite{Fo11} (the significance of the choice $a=1/2 - 1/\beta$ is that the Laguerre ensemble then reduces to the chiral ensemble
with weight $w(\lambda) = e^{-\beta \lambda^2/2}$; see e.g. the discussion in \cite[\S 3.2]{FW11}).
For $\beta = 1,2$ or 4 the validity of (\ref{A1}) can
readily be illustrated numerically by comparing its evaluation against that of the exact expressions for the density known in terms of Laguerre
polynomials (see e.g.~\cite[Ch.~5\& 6]{Fo10}). 
For definiteness consider $\beta = 2$, for which
$$
\rho_{(1)}^{\rm L}(x) = {\Gamma(N+1) \over \Gamma(N+a)} x^a e^{-x} \Big (
L_N^a(x) {d \over dx} L_{N-1}^a(x) - L_{N-1}^a(x) {d \over dx} L_{N}^a(x) \Big )
$$
and $\beta = 1$, $N$ even, for which
\begin{eqnarray*}
\rho_{(1)}^{\rm L}(x)  \Big |_{w(\lambda) = \lambda^{a/2-1} e^{-\lambda/2}}   & = &  \rho_{(1)}^{\rm L}(x)  \Big |_{\beta = 2 \atop
w(\lambda) = \lambda^{a} e^{-\lambda}}  - x^{(a-1)/2} e^{-x/2} L_{N-1}^a(x) \\
&& \times {\Gamma(N) \over 4
\Gamma(a+N-1)} \int_0^\infty {\rm sgn} (x-t) L_{N-2}^a(t) t^{(a-1)/2} e^{-t/2} \, dt.
\end{eqnarray*}
In these expressions set $a = \alpha N$, and denote the corresponding
asymptotic form of  as implied by
(\ref{A1}) by $\rho_{(1)}^{\rm L, \, asym}(Nx)$.  We now perform a numerical computation of the ratio
$$
r_{\beta,N}^{\rm L}(x;\alpha) = { \rho_{(1)}^{\rm L, \, asym}(Nx) \over  \rho_{(1)}^{\rm L}(Nx)}.
$$
The obtained results for given $x$ and varying $N$ are given in Table \ref{ta1}.

\begin{table}
\begin{center}
\begin{tabular}{c||c|c|c||c|c|c}
&0.1 & 6 & 10 & 0.1 & 6 & 10  \\\hline
6 & 1.670 &4.318 &1.122 &  1.072 & 1.205& 1.083 \\
12& 1.357  & 2.978 &  1.062 & 1.049  & 1.101 & 1.041\\
18 &  1.246  &2.460 &  1.041 &  1.039& 1.065 & 1.027\\
24 &  1.189&1.175&  1.031 & 1.027 &1.048 & 1.020\\
30& 1.153 &1.991&1.025 &1.023 &1.039 & 1.016
\end{tabular}
\caption{\label{ta1} The ratios $r^{\rm L}_{2,N}(x;1)$ (first three inner columns) and
$r^{\rm L}_{1,N}(x;1)$ (final three columns) for varying $N$ (rows) and given $x$ (columns).
The eigenvalue support is $(0.171,5.828)$}
\end{center}
\end{table}

\subsection{The Jacobi $\beta$-ensemble}

The derivation of the first non-oscillatory correction term for the Gaussian $\beta$-ensemble density given in \cite[\S 14.3]{Fo10} implies that in the case of the Jacobi $\beta$-ensemble with weight as specified below (\ref{A2}), (\ref{cd3}) can be extended to read
\begin{align}
\nonumber \rho_{(1), N}&(t) = N \frac{2+\alpha_1+\alpha_2}{2\pi} \frac{\sqrt{(t-c_1)(c_2-t)}}{t(1-t)}\\
\label{17.a} &+ \left( \frac{1}{\beta}- \frac{1}{2}\right) \left( \frac{1}{2} \delta (t-c_1) +\frac{1}{2} \delta(t-c_2) -\frac{1}{\pi} \frac{1}{\sqrt{(t-c_1)(c_2-t)}} \right)
+ O \Big ( {1 \over N} \Big )
\end{align}
(cf.~the first asymptotic equality in (\ref{12.a})). 
In a notation analogous to that used in (\ref{11.4A}), to compute $\mu_N^{{\rm J},(1)} - \mu_N^{{\rm J},(2)}$ requires computing (\ref{17.2a}) with
$\rho_{(1),N}(t)$ given by (\ref{17.a}) and $v(t) = \log | x - t|$.

To carry out this task, let the first term in  (\ref{17.a}) be denoted by $N \rho_{(1), \infty}^{\rm J} (x)$. We know from \cite{Wa80} (see also \cite{No11}, \cite{MS11b}) that
\begin{align}
\label{17.b}2 \int_{c_1}^{c_2} \frac{\rho_{(1), \infty}^{\rm J}(x)}{w-x} dx = - \frac{\alpha_1}{w} +\frac{\alpha_2}{1-w} - (\alpha_1+\alpha +2) \frac{u^{\rm J}}{w(1-w)},
\end{align}
where 
\begin{align}\label{17.bx}
u^{\rm J} = \left \{ \begin{array}{ll}
\displaystyle \sqrt{(w - c_1)(w - c_2)}, & w > c_2 \\  \displaystyle
-\sqrt{(w - c_1)(w - c_2)}, &  w < c_1.\end{array}  \right. 
\end{align}
The anti-derivative of the final term can be computed using computer algebra. Choosing the integration constant so that as $w\to \infty$ the constant term vanishes tells us that
\begin{eqnarray}\label{R18}
&& 2 \int_{c_1}^{c_2} \rho_{(1),\infty}^{\rm J}(X) \log | x - X| \, dX  = (\alpha_1 + \alpha_2 + 2) \log \Big | {1 \over 2} \Big (x - {c_1 + c_2 \over 2} + u^{\rm J}
\Big ) \Big | \nonumber \\
&& \qquad \quad +
\alpha_1 \log \Big | {R(x;c_1,c_2) \over x^2 (\sqrt{c_1} - \sqrt{c_2})^2} \Big | +
\alpha_2 \log \Big | {R(1-x;1-c_1,1-c_2) \over (1 - x)^2 (\sqrt{1-c_1} - \sqrt{1-c_2})^2} \Big | 
\end{eqnarray}
where
\begin{equation}\label{R}
R(x;c_1,c_2) := x(c_1 + c_2) - 2c_1 c_2 - 2 \sqrt{c_1 c_2} u^{\rm J}.
\end{equation}
In fact simple manipulation shows the RHS of (\ref{R18}) can alternatively be written
\begin{eqnarray}\label{R1e}
&& 2 \log \Big | {1 \over 2} \Big ( x - {c_1 + c_2 \over 2} + u^{\rm J} \Big ) \Big | \nonumber \\
&& -2 \alpha_1 \log {2 |\sqrt{c_1 c_2} + x - u^{\rm J}| \over (\sqrt{c_1} + \sqrt{ c_2})^2} -
2 \alpha_2 \log {2 |\sqrt{(1 -c_1)(1- c_2)} + 1 -x + u^{\rm J}| \over (\sqrt{1-c_1} + \sqrt{1- c_2)^2}}. 
\end{eqnarray}

The integral required by the second term in (\ref{17.a}) to compute  (\ref{17.2a}) with $v(t) = \log | x - t|$ is given
by (\ref{20.2}). Thus we have
\begin{align}\label{R1}
\mu_N^{{\rm J},(1)} & - \mu_N^{{\rm J},(2)}  =  N \int_{c_1}^{c_2} \rho_{(1),\infty}^{\rm J}(X) \log | x - X| \, dX \nonumber \\
& + \Big ( {1 \over \beta} - {1 \over 2} \Big )
\Big ( {1 \over 2} \log |(x-c_1)(x-c_2)| - \log {1 \over 2} \Big | x - {c_1 + c_2 \over 2} + u^{\rm J} \Big | \Big ) + O\Big ( {1 \over N} \Big ).
\end{align}

We now turn our attention to the calculation of the variance. This task is very similar to that already detailed in the Laguerre case,
and so we will be content with presenting the final result only. Thus we find
\begin{eqnarray}\label{Rs}
(\sigma^2)^{{\rm J},(1)} -  (\sigma^2)^{{\rm J},(2)} & = & - {2 \over \beta} \log |u^{\rm J}| + {2 \over \beta} \log \Big | {1 \over 2} \Big ( x - {c_1 + c_2 \over 2} + u^{\rm J} \Big ) \Big |  - {2 \alpha_1 \over \beta} \log{2  | \sqrt{c_1 c_2} + x - u^{\rm J} | \over (\sqrt{c_1} + \sqrt{c_2})^2}
\nonumber \\
&&- {2 \alpha_2 \over \beta}
\log { 2| \sqrt{(1 - c_1)(1 - c_2)} - (x - 1 - u^{\rm J} )| \over (\sqrt{1 - c_1} + \sqrt{1 - c_2})^2} .
\end{eqnarray}

Substituting (\ref{R1}) and (\ref{Rs}) in (\ref{17.1}) and recalling (\ref{2.11x}) allows us to deduce the asymptotic expansion
\begin{eqnarray}\label{Re}
&&\left\langle \prod_{l=1}^N |x-\lambda_l |^\beta \right\rangle_{{\rm ME}_{\beta, N}(\lambda^{\alpha_1(N+1)\beta/2)} 
(1 - \lambda)^{\alpha_2 (N+1)\beta/2})} \nonumber \\
&& \quad \mathop{\sim}\limits_{N \to \infty}
\exp \Bigg\{  (N+1) \beta \Bigg ( \log \Bigg |{1 \over 2} \Big (
x - {c_1 + c_2 \over 2} + u^{\rm J} \Big ) \Bigg | \nonumber \\
&& \qquad  \quad - \alpha_1 \log {2 | \sqrt{ c_1 c_2} + x - u^{\rm J} | \over (\sqrt{c_1}  +  \sqrt{c_2})^{2}} -
 \alpha_2 \log {2| \sqrt{(1 -c_1)(1 - c_2)} - x + 1 + u^{\rm J}| \over
 (\sqrt{1 - c_1} + \sqrt{1 - c_2})^{2}} \Bigg ) \Bigg \} \nonumber \\
 && \qquad   \quad \times \exp \Bigg \{  (1-3\beta/2) \log  |u^{\rm J} | - (1 - \beta/2) 
\log {1 \over 2} \left | u^{\rm J} +x- (c_1 + c_2)/2 
\right| \Bigg \}. 
\end{eqnarray}
From this, (\ref{2.11}) and the asymptotic formula for the ratio of normalizations
\begin{eqnarray*}
\lefteqn{{C_{\beta,N}(\lambda^{\beta \alpha_1 (N+1)/2 + \beta/2 - 1}(1- \lambda)^{\beta \alpha_2 (N+1)/2 + \beta/2 - 1}) \over
C_{\beta,N+1}(\lambda^{\beta \alpha_1 (N+1)/2 + \beta/2 - 1}(1- \lambda)^{\beta \alpha_2 (N+1)/2 + \beta/2 - 1} )} } \\
&& \sim {\Gamma(1 + \beta / 2) \over 2 \pi} 
\Big ( {2 \over \beta N} \Big )^{\beta / 2}
{((1 + \alpha_1) (1 + \alpha_2) (1 + \alpha_1 + \alpha_2))^{1/2} \over (2 + \alpha_1 + \alpha_2)^{\beta/2 + 1} } \\
&& \quad \times  \exp \Big \{ \beta (N + 1) \Big ( 2 \log (\alpha_1 + \alpha_2 + 2) - {1 \over 2} \log ( \alpha_1 + 1) - {1 \over 2} \log ( \alpha_2 + 1) 
- {1 \over 2} \log (\alpha_1 + \alpha_2 + 1) \Big )\\
&& \quad + \beta (N + 1) \alpha_1 \Big (   \log (\alpha_1 + \alpha_2 + 2) - {1 \over 2} \log ( \alpha_1 + 1) 
- {1 \over 2} \log (\alpha_1 + \alpha_2 + 1) \Big ) \\
&&\quad + \beta (N + 1) \alpha_2 \Big (   \log (\alpha_1 + \alpha_2 + 2) - {1 \over 2} \log ( \alpha_2 + 1) 
- {1 \over 2} \log (\alpha_1 + \alpha_2 + 1) \Big ) \Big \}
\end{eqnarray*}
 (deduced from the Selberg
integral \cite[Ch.~4]{Fo10} and Stirling's formula), and the inter-relations
\begin{align*}
\sqrt{c_2} + \sqrt{c_1} & = {2 (1 + \alpha_1)^{1/2} (1 + \alpha_1+ \alpha_2)^{1/2}  \over (2 + \alpha_1 + \alpha_2)} \\
\sqrt{1-c_2} + \sqrt{1-c_1} & = {2 (1 + \alpha_2)^{1/2} (1 + \alpha_1+ \alpha_2)^{1/2}  \over (2 + \alpha_1 + \alpha_2)} \\
c_2 - c_1 & = {4 (1 + \alpha_1)^{1/2} (1 + \alpha_2)^{1/2} (1 + \alpha_1+ \alpha_2)^{1/2}\over
 (2 + \alpha_1 + \alpha_2)^2}
 \end{align*}
 (deduced from (\ref{cd1}) and (\ref{cd2}))
 we obtain (\ref{A2}).

As for (\ref{A1}), exact formulas for the density at $\beta = 1,2$ and 4 (see e.g.~\cite[Ch.~5\&6]{Fo10}), now involving Jacobi polynomials, can be used
to check  (\ref{A2}). Explicitly, consider the Jacobi weight specified above (\ref{A2}), and let
$\rho_{(1),N}^{\rm J, asym}(x)$ denote the asymptotic density for $x \in (0,c_1) \cup (c_2,1)$.
The ratio
$$
r_{\beta,N}^{\rm J}(x;\alpha_1,\alpha_2) = {\rho_{(1)}^{\rm J, asym}(x) \over \rho_{(1)}^{\rm J}(x)}
$$
can then readily be computed numerically for $\beta 1, 2$ and 4.
As seen in Table \ref{ta2}, results similar to those presented in Table \ref{ta1} are obtained.

\begin{table}
\begin{center}
\begin{tabular}{c||c|c|c||c|c|c}
&0.8 & 0.85 & 0.9 & 0.8 & 0.85 & 0.9  \\\hline
6 & 1.866 & 1.224 &1.115 &  1.045 & 1.055& 1.055 \\
12& 1.487  & 1.116 &  1.058 & 1.030  & 1.040 & 1.028\\
18 &  1.345  &1.079 &  1.038 &  1.033& 1.028 & 1.018\\
24 &  1.269&1.059&  1.029 & 1.035 &1.021 & 1.014\\
30& 1.221 &1.048&1.023 &1.035 &1.017 & 1.011
\end{tabular}
\caption{\label{ta2} The ratios $r^{\rm J}_{2,N}(x;5,5)$ (first three inner columns) and
$r^{\rm J}_{1,N}(x;5,5)$ (final three columns) for varying $N$ (rows) and given $x$ (columns).
The eigenvalue support is $(0.223,0.776)$}
\end{center}
\end{table}

\subsection{Soft edge scaling}
For the Laguerre $\beta$-ensemble with $a = \alpha N$ there is a scaling about the smallest and largest eigenvalues such that the
neighbouring eigenvalues have a spacing $O(1)$. Thus about the largest eigenvalue this occurs in the variable $X$ when \cite{Jo01}
\begin{equation}\label{F1}
N \lambda = N (1 + \sqrt{\alpha + 1})^2 + N^{1/3} (\sqrt{\alpha + 1} +1 ) \Big (1+ {1 \over \sqrt{\alpha + 1}}  \Big )^{1/3} X
\end{equation}
and for the  smallest eigenvalue when \cite{BFP98}
\begin{equation}\label{F2}
N \lambda = N (1 - \sqrt{\alpha + 1})^2 - N^{1/3} (\sqrt{\alpha + 1} -1 ) \Big (1- {1 \over \sqrt{\alpha + 1}}  \Big )^{1/3} X.
\end{equation}
We note that with $t_\pm^{\rm L} := ( \sqrt{\alpha + 1} \pm 1)^2$, these can be written
\begin{equation}\label{3.32'}
\lambda = t_\pm^{\rm L}  \pm N^{-2/3} {(t_\pm^{\rm L})^{2/3} \over ((t_+ - t_-)/4)^{1/3}},
\end{equation}
where the $+(-)$ sign is to be taken about the largest (smallest) eigenvalue.
Moreover, the distributions of the largest and smallest eigenvalues
have well defined limits in terms of $X$, independent of $\alpha$. These are the soft edge distributions,
common --- according to the principle of universality \cite{Ku11} --- to a large class of random matrix ensembles at the spectrum edge,
and in particular the Gaussian $\beta$-ensemble and the (soft edge of) the Laguerre $\beta$-ensemble with parameter $a$ fixed.
For these latter ensembles it is known \cite{DV11,Fo11,BN11} that the distribution of the soft edge scaled largest eigenvalue, $p^{\rm soft}(X)$ say,
has the large $X$ form
\begin{equation}\label{F3}
p^{\rm soft}(X) \mathop{\sim}\limits_{X \to \infty} {1 \over \pi}
{\Gamma(1 + \beta/2) \over (4 \beta)^{\beta/2}} {e^{-2 \beta X^{3/2}/3} \over X^{3 \beta /4 - 1/2}}.
\end{equation}
Replacing $x$ in (\ref{A1}) by the RHS of (\ref{3.32'}), then taking $N \to \infty$ we reclaim (\ref{F3}). 

There is an analogous scaling in the case of the Jacobi $\beta$-ensemble with $a_1 = \alpha_1 N$, $a_2 = \alpha_2 N$.
About the largest ($+$ sign) and smallest ($-$ sign) eigenvalues this reads
\cite{Jo08,Bo11}
\begin{equation}\label{F7}
\lambda = t_{\pm}^{\rm J}  \pm N^{-2/3} {(\tau t_\pm^{\rm J} (1 - t_\pm^{\rm J} ))^{2/3} \over (\tau \theta (1 - \tau ) (1 - \theta))^{1/6}},
\end{equation}
where $t_+^{\rm J}  = c_2$, $t_-^{\rm J} = c_1$ and
$$
\theta := {\alpha_1 + 1 \over \alpha_1 + \alpha_2 + 2}, \qquad \tau := {1 \over  \alpha_1 + \alpha_2 + 2}.
$$
Replacing $x$ in (\ref{A2}) by the RHS of (\ref{F7}) and taking $N \to \infty$ again reclaims (\ref{F3}).

\subsection*{Acknowledgements}
The assistance of
Anthony Mays in the preparation of the manuscript is
acknowledged. This work was supported by the Australian Research Council.

\appendix
\section{Appendix}
\setcounter{equation}{0}

Consider  the Laguerre $\beta$-ensemble with weight $\lambda^{a \beta /2} e^{-\beta \lambda /2}$ and suppose the exponent $a$ is
fixed. The eigenvalue support then extends to the origin, so there is now no region to the left of the interval of support for which the
density is exponentially small. As first noted in \cite{CM94}, the appropriate large deviation quantity is now
$E_{\beta,N}(0;(0,s))$ --- the probability that the interval $(0,s)$ is free of eigenvalues. Computation of the large deviation asymptotics
(using the Dyson log-gas method), setting  $s \mapsto X/4N$, and taking $N \to \infty$ reproduces the $X \to \infty$ asymptotics of
the {\it hard edge} scaled gap probability
$$
E_\beta^{\rm hard}(0;(0,X)) := \lim_{N \to \infty} E_{\beta,N}(0,(0,X/4N)).
$$
According to \cite{Fo94b} (see also the recent works \cite[eq.~(13.52)]{Fo10} and \cite{RRZ11}) the latter reads
\begin{equation}\label{AF1}
\log E_\beta^{\rm hard}(0;(0,X))
\mathop{\sim}\limits_{X \to \infty} - \beta \Big ( {X \over 8} - a {\sqrt{X} \over 2} +
\Big ( {a (a - 1) \over 4} + {a \over 2 \beta} \Big ) \log X^{1/2} +  O(1) \Big )
\end{equation}
(for $a \in \mathbb Z_{\ge 0}$ the explicit form of the constant term is also known; see e.g.~\cite[below eq.~(13.52)]{Fo10}).

There has been a recent work relating to the asymptotic form (\ref{AF1}) \cite{KC10} from the large deviation viewpoint. 
There (eq.~(21) of  \cite{KC10} with $z \in [a^2/4,\infty)$) an expression
functionally simpler than (\ref{AF1}) is presented. However when compared against (\ref{AF1}) it is seen to differ in  the coefficient
of $\log X^{1/2}$, and differ too in the constant term in the cases the latter is known exactly (the large $Nt$ region of the graph in
Figure 2 of \cite{KC10} also reveals this discrepancy). The formula eq.~(21) of  \cite{KC10}  must therefore be regarded as an approximation rather
than an exact result.

%\bibliographystyle{amsplain}
%\bibliography{book1}

\begin{thebibliography}{10}

\bibitem{ATK11}
S.~Adachi, M.~Toda, and H.~Kubotani, \emph{Asymptotic analysis of singular
  values of rectangular complex matrices in the {L}aguerre and fixed trace
  ensembles}, J. Phys. A \textbf{44} (2011), 292002(8pp).

\bibitem{BFP98}
T.H. Baker, P.J. Forrester, and P.A. Pearce, \emph{Random matrix ensembles with
  an effective extensive external charge}, J. Phys. A \textbf{31} (1998),
  6087--6101.

\bibitem{Be97}
C.W.J. Beenakker, \emph{Random-matrix theory of quantum transport}, Rev. Mod.
  Phys. \textbf{69} (1997), 731--808.

\bibitem{Bo11}
F.~Bornemann, \emph{On the scaling limits of determinantal point processes with
  kernels induced by {S}turm-{L}iouville operators}, arXiv:1104.0153, 2011.

\bibitem{BG11}
G.~Borot and A.~Guionnet, \emph{Asymptotic expansion of beta matrix models in
  the one-cut regime}, arXiv:1107.1167, 2011.

\bibitem{BN11}
G.~Borot and C.~Nadal, \emph{Right tail expansion of {T}racy-{W}idom beta
  laws}, arXiv:1111.2761, 2011.

\bibitem{BFB96}
P.W. Brouwer, K.~Frahm, and C.W.J. Beenakker, \emph{Quantum mechanical
  time-delay matrix in chaotic scattering}, Phys. Rev. Lett. \textbf{78}
  (1997), 4737--4740.

\bibitem{CM94}
Y.~Chen and S.M. Manning, \emph{Asymptotic level spacing of the Laguerre
  ensemble: a {C}oulomb fluid approach}, J. Phys. A \textbf{27} (1994),
  3615--3620.

\bibitem{DF06}
P.~Desrosiers and P.J. Forrester, \emph{Hermite and {L}aguerre
  $\beta$-ensembles: asymptotic corrections to the eigenvalue density}, Nucl.
  Phys. B \textbf{743} (2006), 307--332.

\bibitem{DV11}
L.~Dumaz and B.~Vir\'ag, \emph{The right tail exponent of the
  {T}racy-{W}idom-beta distribution}, arXiv:1102.4818, 2011.

\bibitem{DE05}
I.~Dumitriu and A.~Edelman, \emph{Global spectrum fluctuations for the
  $\beta$-{H}ermite and $\beta$-{L}aguerre ensembles via matrix models}, J.
  Math. Phys. \textbf{47} (2006), 063302.

\bibitem{Fo94b}
P.J. Forrester, \emph{Exact results and universal asymptotics in the {Laguerre}
  random matrix ensemble}, J. Math. Phys. \textbf{35} (1994), 2539--2551.

\bibitem{Fo10}
\bysame, \emph{Log-gases and random matrices}, Princeton University Press,
  Princeton, NJ, 2010.

\bibitem{Fo11}
\bysame, \emph{Spectral density asymptotics for {G}aussian and {L}aguerre
  $\beta$-ensembles in the exponentially small region}, arXiv:1111.1350, 2011.

\bibitem{FW11}
P.J. Forrester and N.S. Witte, \emph{Asymptotic forms for hard and soft edge
  general $\beta$ ensembles}, arXiv:1110.4284.

\bibitem{FPNFD11}
M.~Fridman, R.~Pugatch, M.~Nixon, A.A. Friesem, and N.~Davidson,
  \emph{Measuring maximal eigenvalue distribution of {W}ishart random matrices
  with coupled lasers}, arXiv:1012.1282, 2010.

\bibitem{Jo98}
K.~Johansson, \emph{On fluctuation of eigenvalues of random {Hermitian}
  matrices}, Duke Math. J. \textbf{91} (1998), 151--204.

\bibitem{Jo01}
I.M. Johnstone, \emph{On the distribution of the largest principal component},
  Ann. Math. Stat. \textbf{29} (2001), 295--327.

\bibitem{Jo08}
\bysame, \emph{Multivariate analysis and {J}acobi ensembles: {L}argest
  eigenvalue, {T}racy-{W}idom limits and rates of convergence}, Ann. Stat.
  \textbf{36} (2008), 2683--2716.

\bibitem{KC10}
E.~Katzav and I.P. Castillo, \emph{Large deviations of the smallest eigenvalue
  of the {W}ishart-{L}aguerre ensemble}, Phys. Rev. E \textbf{82} (2010),
  040104(R).

\bibitem{Ku11}
A.~Kuijlaars, \emph{Universality}, The {O}xford {H}andbook of {R}andom {M}atrix
  {T}heory (G.~Akemann, J.~Baik, and P.~di~Francesco, eds.), Oxford University
  Press, Oxford, 2011, pp.~103--134.

\bibitem{LV11}
G.~Livan and P.~Vivo, \emph{Moments of {W}ishart-{L}aguerre and {J}acobi
  ensembles of random matrices: application to the quantum transport problem in
  chaotic cavities}, Acta Phys. Pol. B \textbf{42} (2011), 1081--1104.

\bibitem{MV09}
S.N. Majumdar and M.~Vergassola, \emph{Extreme value statistics of eigenvalues
  of {G}aussian random matrices}, Phys. Rev. Lett. \textbf{102} (2009), 060601.

\bibitem{MS11a}
F.~Mezzadri and N.J. Simm, \emph{Moments of the transmission eigenvalues,
  proper delay times and random matrix theory {I}}, J. Math. Phys. \textbf{52} (2011)
  103511(29pp)

\bibitem{MS11b}
\bysame, \emph{Moments of the transmission eigenvalues, proper delay times and
  random matrix theory {II}}, arXiv:1108.2859, 2011.

\bibitem{Mu82}
R.J. Muirhead, \emph{Aspects of multivariate statistical theory}, Wiley, New
  York, 1982.

\bibitem{No11}
M.~Novaes, \emph{Asymptotics of {S}elberg-like integrals by lattice path
  counting}, Ann. Phys. \textbf{326} (2011), 828--838.

\bibitem{PS10}
S.~Pasquetti and R.~Schiappa, \emph{Borel and {S}tokes nonperturbative
  phenomena in topological string theory and $c=1$ matrix models}, Ann. Henri
  Poincar\'e \textbf{11} (2010), 351--431.

\bibitem{RRZ11}
J.A. Ramirez, B.~Rider, and O.~Zeitouni, \emph{Hard edge tail asymptotics},
  Elect.~Comm.~in Probab. \textbf{16} (2011), 741--752.

\bibitem{TV04}
A.M. Tulino and S.~Verd\'u, \emph{Random matrix theory and wireless
  communications}, Foundations and {T}rends in {C}ommuncations and
  {I}nformation {T}heory, vol.~1, Now Publisher, 2004, pp.~1--182.

\bibitem{Wa80}
K.W. Wachter, \emph{The limiting measure of multiple discriminant ratios}, Ann.
  Statist. \textbf{8} (1980), 937--957.

\end{thebibliography}

\providecommand{\bysame}{\leavevmode\hbox to3em{\hrulefill}\thinspace}
\providecommand{\MR}{\relax\ifhmode\unskip\space\fi MR }
\providecommand{\MRhref}[2]{%
  \href{http://www.ams.org/mathscinet-getitem?mr=#1}{#2}
}
\providecommand{\href}[2]{#2}

\end{document}